\documentclass[intlimits,sumlimits,12pt]{iopart}

\newcommand{\diag}{\textrm{diag}}
\newcommand{\dd}{\text{d}}
\newcommand{\trans}{\text{T}}
\newcommand{\ytens}{\left(Y\otimes \mathds{1}_N\right)}
\newcommand{\ytenstrans}{\left(Y^\trans \otimes \mathds{1}_N\right)}

\usepackage{iopams}
\expandafter\let\csname equation*\endcsname\relax

\expandafter\let\csname endequation*\endcsname\relax

\usepackage{amsmath}
\usepackage{hyperref}
\usepackage{dsfont}
\usepackage{cite}
\usepackage{graphicx}
\usepackage{amssymb}
\usepackage{mathtools}
\usepackage[capitalise]{cleveref}
\crefname{section}{Sec.}{Secs.}
\usepackage{braket}
\usepackage[disable]{todonotes}
\usepackage{float}
\usepackage{subfig}
\usepackage{nicematrix}
\NiceMatrixOptions{cell-space-limits = 2pt}

\DeclareMathOperator{\str}{str}
\DeclareMathOperator{\sdet}{sdet}

\newcommand{\tauthree}{\tau^{(3)}}

\newcommand*{\myprime}{^{\prime}\mkern-1.2mu}

\newcommand{\bralign}[1]{\phantom{#1} \quad}
\begin{document}

\title[]{Breaking the Spin Degeneracy for Distributions of Off-Diagonal Scattering Matrix Elements in the Case of Symplectic Symmetry}

\author{Nils Gluth and Thomas Guhr}

\address{Fakult\"at f\"ur Physik, Universit\"at Duisburg--Essen, Duisburg, Germany}
\ead{nils.gluth@uni-due.de, thomas.guhr@uni-due.de}
\vspace{10pt}

\begin{abstract}
  Recently derived exact expressions for the distributions of scattering matrix elements show that all combinations of ingoing and outgoing spin orientations are equally distributed. We propose a model suitable for experimental realization that breaks this degeneracy by introducing a spin preference in the channels. With the help of supersymmetry we derive a corresponding nonlinear sigma model. We exactly calculate the distributions of real and imaginary part in first order of the parameter measuring the preference. Quite remarkably, while breaking the spin degeneracy leads to different distributions for the favored and unfavored spin orientation, the real and imaginary parts of these altered distributions are the same as in the case of unbroken spin degeneracy.
\end{abstract}

\vspace{2pc}

\submitto{}

\section{Introduction}
\label{sec0}
In a recent work the authors of Ref. \cite{HSSK2025} managed to observe spin resonances in a microwave network. It realizes a symplectic system with modulated bond lengths which introduce phases dependent on the spin orientation. These systems in themselves do not possess a spin and obtain the GSE features through geometric effects. As shown in Ref. \cite{HSSK2025} observation of spin dependent effects are still possible. This among other recent experiments \cite{LCZD2020,CLZDC2021,LAFBBDS2023,CGKGD2025} show that there is an ever growing interest in GSE systems and the observation of spin effects. Our recent works \cite{GG2025} showed that in genuine GSE systems the scattering elements for different spin orientations are distributed equally due to a present spin orientation symmetry. This motivates us to put forward a model capable of breaking the aforementioned spin orientation symmetry and allows observation of universal effects dependent on the spin orientation. Furthermore, experimental applications of this model are realizable by minor modifications of GSE graphs suitable for measuring off-diagonal scattering matrix elements, for example by adding attenuation to the antennas.
This paper is the second of a series of three papers. We refer the reader to \cite{GG2025} for an introduction to the topic and some of the salient features and the literature.

The paper is structured as follows. We discuss the possible ways in which we can introduce a spin preference to the system at hand and justify our approach in \cref{sec:breakspinsym}. We exactly calculate the distribution of real and imaginary part of the scattering matrix elements via supersymmetry in \cref{sec:calcdistr}. Finally, we discuss our results in \cref{sec:conclusion}

\section{Breaking of Spin Orientation Symmetry}
\label{sec:breakspinsym}
In recent studies \cite{GG2025} we derived an explicit expression for the distribution of the off-diagonal scattering matrix element for the Gaussian Symplectic Ensemble using supersymmetry. One of the main observations was that for all four entries of the matrix representation of the quaternion scattering element $S_{ab}$ not only their respective real and imaginary parts are equally distributed but also all four entries of $S_{ab}$ for the possible spin orientations are equally distributed. This is so because both the GSE Hamiltonian $H$ and the coupling vectors $W_c$ do not distinguish the spin orientations. The system not only has a spin inversion symmetry but actually does not discern between the spin states in the channels at all. Here, we aim to break up this spin degeneracy, which is of relevance for experimental applications. 

The coincidence of the distribution with respect to the spin orientations arises from the symplectic structure of the Hamiltonian as well as of the coupling vectors. There are two ways to break up this degeneracy. The first option is adding a perturbation $\alpha H_0$ to the Hamiltonian to break the symplectic symmetry where $\alpha$ will mediate how strongly this symmetry is broken. However, this brings about a few issues as it is not intuitively clear what properties, if any, $H_0$ should have and if it is fixed or itself distributed according to some ensemble. Depending on the choice of $H_0$, the supersymmetric formalism drastically changes and thus also the resulting non-linear sigma model. The experimental realizations of such systems as microwave networks are already difficult for symplectic graphs \cite{CGKGD2025} and are even more challenging with additional constraints \cite{HSSK2025}.

The second option is to break the spin orientation symmetry of the coupling vectors $W_c$. This has the advantage that experimental realizations are simple extensions of the setups used to measure GSE graphs. For example the addition of attenuators to certain antennas break the spin orientation symmetry of the channels. Other realizations of GSE systems, as put forward in Ref. \cite{KBSR2008}, may also be modified by adding external magnetic fields to the channels to induce splitting depended on the spin orientation. In general additional constraints on the channels are easier realized than constraints on the system itself. Furthermore, breaking the spin orientation symmetry in the channels has the further benefit that much of the formalism derived in \cite{GG2025} is still applicable. To that end we introduce a preference for one of the two spin directions by modifying the coupling vectors via
\begin{equation}\label{eqn:wceps}
	W_c = \begin{bNiceMatrix}
		\left(1 + \varepsilon\right) \widetilde{W}_c & \left(1 - \varepsilon\right) \ytens \widetilde{W}_c^\star
	\end{bNiceMatrix}
\end{equation}
where the parameter $\varepsilon\in[0,1]$ controls how strongly one of the spins is favored. Consequently, changing the coupling vectors also changes the scattering matrix elements
\begin{equation}
	S_{ab} = - i 2\pi \begin{bNiceMatrix}
		\left(1 + \varepsilon\right)^2 \widetilde{W}_a^\dagger G \widetilde{W}_b & \left(1-\varepsilon^2\right) \widetilde{W}_a^\dagger G \ytens \widetilde{W}_b^\star \\
		\left(1 - \varepsilon^2\right) \widetilde{W}_a^\trans \ytenstrans G \widetilde{W}_b & \left(1 - \varepsilon\right)^2 \widetilde{W}_a^\trans \ytenstrans G \ytens \widetilde{W}_b^\star 
	\end{bNiceMatrix} 
\end{equation}
where the matrix resolvent $G$ also contains the modified coupling vectors, see the definitions in Ref. \cite{GG2025}.

\section{Deriving the Distribution of Off-Diagonal Scattering Elements}
\label{sec:calcdistr}
Here we only present an overview of the steps necessary for calculating the distribution via supersymmetry. The details follow in Ref. \cite{GG2025}.We only focus on the necessary changes needed to calculate the distribution which arise from breaking the spin symmetry of the coupling vectors $W_c$. 

In \cref{subsec:Distribution}, we introduce the distribution and characteristic function of the off-diagonal scattering matrix elements. In \cref{subsec:HubStrTransformation}, we map the characteristic function on superspace and subsequently carry out the integration over a real supervector in \cref{subsec:RealSupervector}. Then, we carry out a saddle point approximation in \cref{subsec:SaddlePoint} and integrate over the resulting saddle point manifold in \cref{subsec:IntegralSaddlePoint}. Concluding this section, we discuss regime of validity of our approach in \cref{subsec:Validity}
\subsection{Distribution of Off-Diagonal Scattering Matrix elements as an Ensemble Average}
\label{subsec:Distribution}
The distribution is the ensemble average of a filter function
\begin{equation}
	P_{s,m m\myprime}\left(x_s\right) = \int\dd[H] \mathcal{P}^{(4)}\left(H\right) \delta\left(x_s - \wp_s \left(S_{am bm\myprime}\left(E\right)\right)\right)
\end{equation}
where the distribution $\mathcal{P}^{(4)}(H)$ and the volume element $\dd[H]$ are defined in Eq. (9) and (11) of Ref. \cite{GG2025}, respectively, and the real and imaginary part
\begin{equation}
	\wp_s\left(S_{am bm\myprime}\left(E\right)\right) = \frac{1}{2 i^{s-1}} \left(S_{am bm\myprime}\left(E\right) + (-1)^{s-1} S_{am bm\myprime}^\star\left(E\right)\right)
\end{equation}
with the spin orientations $m, m\myprime\in\{\uparrow, \downarrow\}$. The associated characteristic function is
\begin{equation}
	R_{s,m m\myprime}(k) = \int\dd[H] \mathcal{P}^{(4)}(H) \exp\left(- i k \wp_s\left(S_{am bm\myprime}\left(E\right)\right)\right)
\end{equation}
and we write the real and imaginary part as a quadratic form 
\begin{equation}
	\wp_s\left(S_{am bm\myprime}(E)\right) = \pi \widetilde{W}^\dagger A_s \widetilde{W} \quad \text{with} \quad A_s = \begin{bNiceMatrix}
		0 & (-i)^s G \\
		i^s G^\dagger & 0
	\end{bNiceMatrix}
\end{equation}
with the vectors
\begin{align}
	\widetilde{W}_{\uparrow\uparrow} =& \left(\left(1 + \varepsilon\right)\widetilde{W}_a, \left(1 + \varepsilon\right)\widetilde{W}_b\right) \notag\\
	\widetilde{W}_{\uparrow\downarrow} =& \left(\left(1 + \varepsilon\right)\widetilde{W}_a, \left(1 - \varepsilon\right) \ytens \widetilde{W}_b^\star\right) \notag\\
	\widetilde{W}_{\downarrow\uparrow} =& \left(\left(1 - \varepsilon\right) \ytens\widetilde{W}_a^\star, \left(1 + \varepsilon\right)\widetilde{W}_b\right) \notag\\
	\widetilde{W}_{\uparrow\uparrow} =& \left(\left(1 - \varepsilon\right) \ytens \widetilde{W}_a^\star, \left(1 - \varepsilon\right) \ytens \widetilde{W}_b^\star\right) .
\end{align}
Replacing the phase factor by integrals over the $8N$-dimensional supervector $\psi=(z,\zeta)$ consisting of vectors of commuting $z=(z_a, z_b)$ and anticommuting variables $\zeta = (\zeta_a, \zeta_b)$
\begin{align}
	R_{s,mm\myprime}(k) = \int\dd[\psi] \exp\left(\frac{i}{2} \left(\psi^\dagger \widetilde{\mathbf{W}} + \widetilde{\mathbf{W}}^\dagger \psi \right)\right) \int\dd[H] \mathcal{P}^{(4)}(H) \exp\left(\frac{i}{4\pi k} \psi^\dagger \mathbf{A}_s^{-1} \psi\right)
\end{align}
with $\widetilde{\mathbf{W}}= (\widetilde{W}, 0)$ where $0$ is the $4N$-dimensional zero vector and $\mathbf{A}_s^{-1} = \mathds{1}_2 \otimes A_s^{-1}$. Taking advantage of the fact that $z$ and $z^\dagger$ are independent variables we block diagonalize the matrix $\mathbf{A}_s^{-1}$ by a transformation of $z$ and $\zeta$ and find
\begin{align}
	R_{s, mm\myprime}(k) =& \int\dd[\psi] \exp\left(\frac{i}{2} \left(\psi^\dagger \widetilde{\mathbf{W}} + \widetilde{\mathbf{U}}_s^\dagger \psi\right)\right) \exp\left(\frac{i}{4\pi k} \psi^\dagger \mathbf{\Omega}_0 \psi\right) \notag\\
	&\times \int\dd[H] \mathcal{P}^{(4)}(H) \exp\left(\frac{i}{4\pi k} \tr H K\right) \\
	\widetilde{\mathbf{U}}_s^\dagger =& \widetilde{\mathbf{W}}^\dagger \left(\Xi^+ \oplus \Xi^-\right), \quad \Xi^\pm = \begin{bNiceMatrix}
		0 & \pm\left(-i\right)^s \mathds{1}_{2N} \\
		-i^s \mathds{1}_{2N} & 0
	\end{bNiceMatrix}, \notag\\
	K =& z_a z_a^\dagger - z_b z_b^\dagger - \zeta_a \zeta_a^\dagger - \zeta_b \zeta_b^\dagger, \notag\\
	\mathbf{\Omega}_0 =& E \diag\left(-1,1,-1,-1\right) \otimes\mathds{1}_{2N} + i \pi \diag\left(1,1,1,-1\right) \otimes \sum_{c=1}^{M} W_c W_c^\dagger . 
\end{align}
\subsection{Hubbard-Stratonovich-Transformation}
\label{subsec:HubStrTransformation}
The integration over $H$ yields a Gaussian-like function
\begin{equation}
	\int\dd[H] \mathcal{P}^{(4)}(H) \exp\left(\frac{i}{4\pi k}\tr H K\right) = \exp\left(- \frac{v^2}{2N \left(8\pi k\right)^2} \tr \widehat{K}^2\right)
\end{equation}
of the matrix
\begin{equation}
	\widehat{K} = K + \ytenstrans K^\trans \ytens .
\end{equation}
Due to the duality between the trace and supertrace we replace the matrix containing commuting and anticommuting variables $\widehat{K}$ by a supermatrix $B$
\begin{equation}
	\tr \widehat{K}^2 = \str B^2 .
\end{equation}
We employ another Fourier transform, in this case in superspace,
\begin{equation}
	\exp\left(- \frac{v^2}{2N \left(8\pi k\right)^2} \str B^2\right) = \mathcal{N} \int\dd[\sigma] \exp\left(- \frac{N \left(8\pi k\right)^2}{2v^2} \str\sigma^2\right) \exp\left(i \str\sigma B\right)
\end{equation}
involving a proper supermatrix $\sigma$. Here, $\sigma$ inherits the symmetries of $B$ as guaranteed by
\begin{align}
	\sigma =& T^{-1} \sigma_D T, \quad \sigma_D = \mathfrak{V} \diag\left(\sigma_{B,1} \mathds{1}_2, \sigma_{B,2} \mathds{1}_2, i \sigma_{F,1}, i \sigma_{F,2}, i \sigma_{F,3}, i \sigma_{F,4}\right) \mathfrak{V}^\dagger, \notag\\
	\mathfrak{V} =& \mathds{1}_2 \otimes \diag\left(\mathds{1}_2, \mathfrak{v}\right), \quad \mathfrak{v} = \frac{1}{\sqrt{2}} \begin{bNiceMatrix}
		1 & i \\
		1 & -i
	\end{bNiceMatrix} .
\end{align}
The transformations $T$ preserving the symmetries of $B$ belong to the non-compact unitary orthosymplectic supergroup $\text{UOSp}(2,2\vert 4)$ fulfilling
\begin{align}
	T^\dagger \widetilde{L} T =& \widetilde{L}, \quad \widetilde{L} = \diag\left(1,-1,1,1\right) \otimes\mathds{1}_2, \notag\\
	C T^\star C^\trans =& T, \quad C = \diag\left(Y, Y^\trans, X, X\right), X = \begin{bNiceMatrix}
		0 & 1 \\
		1 & 0
	\end{bNiceMatrix} .
\end{align}

\subsection{Integration over Real Supervectors}
\label{subsec:RealSupervector}
We now carry out the integration over the supervector $\psi$ and it turns out, $\str\sigma B$ is equal to a quadratic form
\begin{equation}
	\str \sigma B = \Psi^\dagger \left(\widetilde{L}^{1/2} \sigma \widetilde{L}^{1/2} \otimes \mathds{1}_{2N}\right) \Psi
\end{equation}
with the $16N$-dimensional supervector 
\begin{equation}
	\Psi = \left(z_a, \ytens z_a^\star, z_b, \ytens z_b^\star, \zeta_a, \ytens \zeta_a^\star, \zeta_b, \ytens \zeta_b^\star\right) .
\end{equation}
Unfortunately, it is not possible to express this quadratic form through the supervector $\psi$ and we are forced to rewrite the rest of the integral in terms of $\Psi$ as well. To that end we have
\begin{equation}
	\widetilde{\mathbf{U}}_s^\dagger \psi + \psi^\dagger \widetilde{\mathbf{W}} = \widetilde{\mathbf{V}}_s^\trans \left(\mathds{1}_4 \otimes \diag\left(\mathds{1}_{2N}, \ytenstrans\right)\right) \Psi, \quad \widetilde{\mathbf{V}}_s^\trans = \begin{bNiceMatrix}
		\widetilde{V}_s^\trans & 0
	\end{bNiceMatrix}
\end{equation}
with the vectors 
\begin{align}
	\widetilde{V}_{s,\uparrow\uparrow} =& \left(1 + \varepsilon\right)\left(
		-i^s  \widetilde{W}_b^\star, \widetilde{W}_a, (-i)^s \widetilde{W}_a^\star, \widetilde{W}_b\right) \notag\\
	\widetilde{V}_{s,\uparrow\downarrow} =& \left(
		-i^s \left(1 - \varepsilon\right) \ytens\widetilde{W}_b, \left(1 + \varepsilon\right) \widetilde{W}_a, (-i)^s \left(1 + \varepsilon\right) \widetilde{W}_a^\star, \left(1 - \varepsilon\right) \ytens \widetilde{W}_b^\star 
	\right)\notag\\
	\widetilde{V}_{s,\downarrow\uparrow} =& \left(
		-i^s \left(1 + \varepsilon\right) \widetilde{W}_b^\star, \left(1 - \varepsilon\right) \ytens\widetilde{W}_a^\star, (-i)^s \left(1 - \varepsilon\right) \ytens\widetilde{W}_a, \left(1 + \varepsilon\right) \widetilde{W}_b
	\right) \notag\\
	\widetilde{V}_{s,\downarrow\downarrow} =& \left(1 - \varepsilon\right) \left(\mathds{1}_4 \otimes \ytens\right)\left(
		-i^s  \widetilde{W}_b, \widetilde{W}_a^\star, (-i)^s  \widetilde{W}_a, \widetilde{W}_b^\star
	\right) .
\end{align}
However, as we have broken the symplectic symmetry of $W_c W_c^\dagger$ by introducing a spin preference we have
\begin{align}
	W_c W_c^\dagger =& \left(1 + \varepsilon^2\right) \left(\widetilde{W}_c \widetilde{W}_c^\dagger + \ytens \widetilde{W}_c^\star \widetilde{W}_c^\trans \ytenstrans\right) \notag\\
	&+ 2 \varepsilon \left(\widetilde{W}_c \widetilde{W}_c^\dagger - \ytens \widetilde{W}_c^\star \widetilde{W}_c^\trans \ytenstrans\right)
\end{align}
which does not have the symplectic symmetry which is present for $\varepsilon=0$. Fortunately, we find a similar relation
\begin{align}
	&z^\trans \ytenstrans W_c W_c^\dagger \ytens z^\star \notag\\
	=& z^\dagger \Bigl\{ \left(1 + \varepsilon^2\right) \left(\widetilde{W}_c \widetilde{W}_c^\dagger + \ytens \widetilde{W}_c^\star \widetilde{W}_c^\trans \ytenstrans\right) \notag\\
	&\phantom{z^\dagger}- 2 \varepsilon \left(\widetilde{W}_c \widetilde{W}_c^\dagger - \ytens \widetilde{W}_c^\star \widetilde{W}_c^\trans \ytenstrans\right) \Bigr\} z
\end{align}
and equivalently for the anticommuting variables. Importantly, we observe that the expression is not invariant but instead the term linear in $\varepsilon$ changes its sign. Expressing $\psi$ through the larger supervector $\Psi$ we have
\begin{align}
	2 \psi^\dagger \mathbf{\Omega}_0 \psi &= \Psi^\dagger \Bigl\{
		- E \widetilde{L} \otimes\mathds{1}_{2N} + i \pi \widetilde{L} L \otimes \left(1 + \varepsilon^2\right) \left(\widetilde{W}_c \widetilde{W}_c^\dagger + \ytens \widetilde{W}_c^\star \widetilde{W}_c^\trans \ytenstrans\right) \notag\\
		&\bralign{\Psi^\dagger}+ i \pi \widetilde{L} L \left(\mathds{1}_4 \otimes \tau^{(3)}\right) \otimes 2 \varepsilon \left(\widetilde{W}_c \widetilde{W}_c^\dagger - \ytens \widetilde{W}_c^\star \widetilde{W}_c^\trans \ytenstrans\right)
	\Bigr\} \Psi, \notag\\
	L &= \diag\left(1,-1,1,-1\right) \otimes \mathds{1}_2 .
\end{align}
where the sign change is reflected in the additional factor containing the third Pauli matrix $\tau^{(3)}$. Hence, the characteristic function is 
\begin{align}\label{eqn:cfSigma}
	R_{s, m m\myprime}(k) =& \mathcal{N} \int\dd[\sigma] \exp\left(- \frac{N (8\pi k)^2}{2v^2} \str\sigma^2\right) \notag\\
	&\times \int\dd[\Psi] \exp\left(\frac{i}{2} \widetilde{\mathbf{V}}_s^\trans \left(\mathds{1}_4 \otimes \diag\left(\mathds{1}_{2N},\ytenstrans\right)\right) \Psi\right) \notag\\
	&\times \exp\left(\Psi^\dagger \left(\widetilde{L}^{1/2} \otimes \mathds{1}_{2N}\right) \Sigma \left(\widetilde{L}^{1/2} \otimes \mathds{1}_{2N}\right) \Psi\right), \notag\\
	\Sigma =& \sigma_E \otimes \mathds{1}_{2N} + \frac{i}{8k} \sum_{c=1}^{M} \left(1 + \varepsilon^2\right) L \otimes \left(\widetilde{W}_c \widetilde{W}_c^\dagger + \ytens \widetilde{W}_c^\star \widetilde{W}_c^\trans \ytenstrans\right) \notag\\
	&+ \frac{i}{8k} \sum_{c=1}^{M} 2 \varepsilon L \left(\mathds{1}_4 \otimes \tauthree\right) \otimes \left(\widetilde{W}_c \widetilde{W}_c^\dagger - \ytens \widetilde{W}_c^\star \widetilde{W}_c^\trans \ytenstrans\right), \notag\\
	\sigma_E =& \sigma - \frac{E}{8\pi k} \mathds{1}_8 .
\end{align}
Transforming from the complex supervector $\Psi$ to a real set of variables $\Phi = \left(x_a, y_a, x_b, y_b, \zeta_a, \zeta_a^\star, \zeta_b, \zeta_b^\star\right)$ via
\begin{equation}
	\Psi = D \Phi, \quad D = \left(\mathds{1}_4 \otimes \diag\left(\mathds{1}_{2N}, \ytens\right)\right) \left(\diag\left(\mathfrak{v},\mathfrak{v},\mathds{1}_4\right) \otimes \mathds{1}_{2N}\right), \quad \mathfrak{v} = \frac{1}{2} \begin{bNiceMatrix}
		1 & i \\
		1 & -i
	\end{bNiceMatrix}
\end{equation}
which does not carry a Berizinian as we use scaled real and imaginary parts. Following \cite{GG2025} we require that the supermatrix
\begin{equation}
	\widetilde{\Sigma} = D^\dagger \left(\widetilde{L}^{1/2} \otimes \mathds{1}_{2N}\right) \Sigma \left(\widetilde{L}^{1/2} \otimes \mathds{1}_{2N}\right) D
\end{equation}
fulfills
\begin{align}\label{eqn:symmetriesSigma}
	\widetilde{\Sigma}_{\text{BB}} =& \widetilde{\Sigma}_{\text{BB}}, \quad J_2 \widetilde{\Sigma}_{\text{FF}} = - \left(J_2 \widetilde{\Sigma}_{\text{FF}}\right)^\trans \quad \text{and} \quad \widetilde{\Sigma}_{\text{FB}}^\trans = - \widetilde{\Sigma}_{\text{BF}} J_2, \notag\\ 
	J_2 =& \diag\left(Y \otimes \mathds{1}_{2N}, Y \otimes \mathds{1}_{2N}\right)
\end{align}
to carry out the integration. The conditions are trivially fulfilled for the first two terms of $\Sigma$ in \cref{eqn:cfSigma} as they are the same as in the symplectic case. Explicit calculations show that the third term also fulfills \cref{eqn:symmetriesSigma} and we arrive at
\begin{align}\label{eqn:cfBeforeSaddlePoint}
	R_{s, mm\myprime}(k) =& \mathcal{N} \int\dd[\sigma] \exp\left(-\frac{N \left(8\pi k\right)^2}{2v^2} \str\sigma^2\right) \sdet^{-1/2}\Sigma \exp\left(- \frac{i}{16} F_{s,m m\myprime}\left(\varepsilon\right)\right), \notag\\
	F_{s,m m\myprime}\left(\varepsilon\right) =& \widehat{\mathbf{V}}_s^\trans \left(\widetilde{L}^{1/2} \otimes \mathds{1}_{2N}\right) \Sigma^{-1} \left(\widetilde{L}^{1/2} \otimes \mathds{1}_{2N}\right) \overline{\mathbf{V}}_s, \notag\\
	\widehat{\mathbf{V}}_s^\trans =& \widetilde{\mathbf{V}}_s^\trans \left(\mathds{1}_4 \otimes \diag\left(\mathds{1}_{2N}, \ytenstrans\right)\right), \notag\\
	\overline{\mathbf{V}}_s =& \left(\mathds{1}_4 \otimes \diag\left(\mathds{1}_{2N}, \ytens\right)\right) \left(\diag\left(X,X,\mathds{1}_4\right)\otimes\mathds{1}_{2N}\right) \widetilde{\mathbf{V}}_s .
\end{align}
Due to the broken symplectic symmetry the inverse $\Sigma^{-1}$ is slightly altered
\begin{align}\label{eqn:SigmaInverse}
	\Sigma^{-1} =& \sigma_E^{-1} \otimes \mathds{1}_{2N} - \sigma_E^{-1} \otimes \sum_{c=1}^{M} \frac{\pi}{\gamma_c} \left(\widetilde{W}_c \widetilde{W}_c^\dagger + \ytens \widetilde{W}_c^\star \widetilde{W}_c^\trans \ytenstrans\right) \notag\\
	&+ \sum_{c=1}^{M} \rho_+^{(c)} \otimes \frac{\pi}{\gamma_c} \widetilde{W}_c \widetilde{W}_c^\dagger + \sum_{c=1}^{M} \rho_-^{(c)} \otimes \frac{\pi}{\gamma_c} \ytens \widetilde{W}_c^\star \widetilde{W}_c^\trans \ytenstrans, \notag\\
	\rho_\pm^{(c)} =& \left(\sigma_E + \frac{i \gamma_c}{8\pi k } L_\pm(\varepsilon)\right)^{-1}, \quad L_\pm(\varepsilon) =  \left(1 + \varepsilon^2\right) L \pm 2 \varepsilon L \left( \mathds{1}_4 \otimes \tauthree\right) .
\end{align}
Thus, after explicit calculations the phase is
\begin{align}\label{eqn:fsdef}
	F_{s, \uparrow\uparrow}(\varepsilon) =& \left(1 + \varepsilon\right)^2 (-i)^{s+1} \frac{\gamma_a}{\pi} \left(\rho_{+,31}^{(a)} + \rho_{-,24}^{(a)}\right) + \left(1 + \varepsilon\right)^2 i^{s+1} \frac{\gamma_a}{\pi} \left(\rho_{+,13}^{(b)} + \rho_{-,42}^{(b)}\right) \notag\\
	F_{s, \uparrow\downarrow}(\varepsilon) =& \left(1 + \varepsilon\right)^2 (-i)^{s+1} \frac{\gamma_a}{\pi} \left(\rho_{+,31}^{(a)} + \rho_{-,24}^{(a)}\right) + \left(1 - \varepsilon\right)^2 i^{s+1} \frac{\gamma_a}{\pi} \left(\rho_{-,13}^{(b)} + \rho_{+,42}^{(b)}\right) \notag\\
	F_{s, \downarrow\uparrow}(\varepsilon) =& \left(1 - \varepsilon\right)^2 (-i)^{s+1} \frac{\gamma_a}{\pi} \left(\rho_{-,31}^{(a)} + \rho_{+,24}^{(a)}\right) + \left(1 + \varepsilon\right)^2 i^{s+1} \frac{\gamma_a}{\pi} \left(\rho_{+,13}^{(b)} + \rho_{-,42}^{(b)}\right) \notag\\
	F_{s, \downarrow\downarrow}(\varepsilon) =& \left(1 - \varepsilon\right)^2 (-i)^{s+1} \frac{\gamma_a}{\pi} \left(\rho_{-,31}^{(a)} + \rho_{+,24}^{(a)}\right) + \left(1 - \varepsilon\right)^2 i^{s+1} \frac{\gamma_a}{\pi} \left(\rho_{-,13}^{(b)} + \rho_{+,42}^{(b)}\right) 
\end{align}
where we observe that the phase does explicitly depend on the spin orientations $m$ and $m\myprime$. Furthermore, the spin orientations determine the sign of $\varepsilon$ for the channels $a, b$ but we stress that this dependence is non-trivial as $\rho_\pm^{(c)}$ depends reciprocally on the metric $L_\pm(\varepsilon)$, \textit{cf.} \cref{eqn:SigmaInverse}. Reassuringly, the limit $\varepsilon\to 0$ recovers the symplectic case and the phases become degenerate to the spin orientations again.
\subsection{Saddle Point Approximation}
\label{subsec:SaddlePoint}
We proceed with a saddle point approximation of \cref{eqn:cfBeforeSaddlePoint} in the same fashion as we did in \cite{GG2025}. The changes in the phase, $F_s(\varepsilon)$ instead of $F_s^{(4)}$, are not relevant for the saddle point approximation as they do not contribute to the saddle point equation and are only evaluated at the saddle point itself. We obtain the non-linear sigma model
\begin{align}\label{eqn:cfnonlinearsigma}
	R_{s, mm\myprime}(k) =& \mathcal{N} \int\dd\mu(Q) \exp\left(- \frac{i}{16} F_{s,m m\myprime}\left(\varepsilon\right)\right) \prod_{c=1}^{M} \sdet^{-1}\left(\mathds{1}_8 + \frac{i \gamma_c}{8\pi k} \sigma_{G,E}^{-1} L\right), \notag\\
	\sigma_G =& \frac{E}{16 \pi k} \mathds{1}_8 - \frac{ \Delta}{16\pi k} Q, \quad \Delta = \sqrt{4v^2 - E^2}
\end{align}
with $Q \in \text{UOSp}(2,2\vert 4)/ \text{UOSp}(2\vert 2)\times \text{UOSp}(2\vert 2)$ from the saddle-point manifold. As it is clear that from this point onward all calculations are performed at the saddle-point we drop the additional index $G$. Next, we expand the exponential containing $F_{s,m m\myprime}(\varepsilon)$ in the anticommuting variables parametrizing $Q$. We use the parametrization derived in \cite{GG2025}, see \ref{app:parametrization} for details, such that
\begin{align}\label{eqn:rhopm}
	\rho_\pm^{(c)} =& - \frac{4\pi k}{v^2} \mathcal{U}^{-1} d_\pm(\varepsilon) \left(\mathds{1}_8 + \frac{i \gamma_c}{2v^2} \widehat{L}_\pm(\varepsilon) \left(E \mathds{1}_8 - i \Delta Q_0\right)\right) \mathcal{U} \notag\\
	d_\pm(\varepsilon) =& \frac{v^2}{\gamma_c \Delta} \left(g_c^+ \mathds{1}_8 + \frac{1}{2} \left(\widehat{L}_\pm(\varepsilon) Q_0 + Q_0 \widehat{L}_\pm(\varepsilon)\right)\right)^{-1}, \notag\\ \widehat{L}_\pm(\varepsilon) =& \mathcal{U} L_\pm(\varepsilon) \mathcal{U}^{-1}, \quad g_c = \frac{\gamma_c^2 + v^2}{\gamma_c \Delta} .
\end{align}
Unlike in the symplectic case $d_\pm(\varepsilon)$ is not diagonal up to a unitary transformation because $[\mathcal{U},\widehat{L}_\pm(\varepsilon)]\neq 0$ and there does not seem to exist an explicit expression for the inverse. Hence, we turn to an expansion of the inverse in powers of $\varepsilon d_1$ where
\begin{equation}
	d_1^{-1} = g_c \mathds{1}_4 + \diag\left(\cosh\theta\mathds{1}_2, \cos\left(\theta_1 + \theta_2\right), \cos\left(\theta_1 - \theta_2\right)\right)
\end{equation}
which effectively is an approximation $\varepsilon/g_c$ for all but one case which we will discuss at the end of our calculations. Hence, we require that $\varepsilon$ is small compared to the partial widths $\gamma_c$. In this approximation we find in first order
\begin{align}
	d_\pm(\varepsilon) \simeq& \frac{v^2}{\Delta \gamma_c} \left(g_c \mathds{1}_8 + \frac{1}{2} \left[L,Q_0\right]_+\right)^{-1} \notag\\
	&\times \left(\mathds{1}_8 \mp \varepsilon \left(g_c \mathds{1}_8 + \frac{1}{2} \left[L,Q_0\right]_+\right)^{-1} \left[L \mathcal{U} \left(\mathds{1}_4 \otimes \tauthree\right) \mathcal{U}^{-1}, Q_0\right]_+\right)
\end{align}
and consequently in $[1,2]$-block notation
\begin{align}\label{eqn:rhopmelem}
	\rho_\pm^{(c)} =& - \frac{4\pi k}{\Delta \gamma_c} \mathcal{U}^{-1} \mathfrak{V} \Biggl\{
		\left(\mathds{1}_2 \otimes d_1\right) \left( \left(E \mathds{1}_8 + i \Delta Q_0\myprime\right) + i 2\gamma_c \left(L \pm 2 \varepsilon L \mathfrak{V}^\dagger \left(\mathds{1}_4 \otimes \tauthree\right) \mathfrak{V}\right)\right) \notag\\
		&\mp \varepsilon \left(\mathds{1}_2 \otimes d_1^2\right) \left[L \mathfrak{V}^\dagger \mathcal{U} \left(\mathds{1}_4 \otimes \tauthree\right) \mathcal{U}^{-1} \mathfrak{V}, Q_0\myprime\right]_+ \left(E \mathds{1}_8 + i \Delta Q_0\myprime + i 2 \gamma_c L\right)
	\Biggr\} \mathfrak{V}^\dagger \mathcal{U} ,\notag\\
	\mathfrak{V} =& \diag\left(\mathds{1}_2, \mathfrak{v}, \mathds{1}_2, \mathfrak{v}\right), \notag\\
	Q_0\myprime =& \begin{bNiceMatrix}
		\cos\widetilde{\theta} & \sin\widetilde{\theta} \\
		\sin\widetilde{\theta} & -\cos\widetilde{\theta}
	\end{bNiceMatrix}, \quad \widetilde{\theta} = \diag\left(\theta \mathds{1}_2, \theta_1 + \theta_2, \theta_1 - \theta_2\right) .
\end{align}
Not surprisingly, we find that $\rho_\pm^{(c)}$ splits into a symplectic part $\rho^{(c)}_\text{GSE}$ independent of $\varepsilon$ and a symmetry breaking part $\rho^{(c)}_{\pm,\text{SB}}$ 
\begin{equation}
	\rho_\pm^{(c)} = \rho^{(c)}_\text{GSE} + \varepsilon \rho^{(c)}_{\pm,\text{SB}} .
\end{equation}
The resulting phase $F_{s,m m\myprime}$ is in first order in $\varepsilon$
\begin{equation}
	F_{s,m m\myprime} = F_s^{(4)} + \varepsilon F_{s,m m\myprime}^{\text{SB}}
\end{equation}
and we expand the exponential in $\varepsilon$ up to linear order such that characteristic function also splits into two parts
\begin{equation}\label{eqn:cfsplit}
	R_{s,m m\myprime}(k) = R^{\text{GSE}}(k) + \varepsilon R_{s,m m\myprime}^{\text{SB}}(k) 
\end{equation}
where $R^{\text{GSE}}(k)$ is the characteristic function obtained in the symplectic case and $R_{m m\myprime}^{\text{SB}}(k)$ contains an additional $F_{s,m m\myprime}^{\text{SB}}$ inside the integral. Not expanding the exponential in terms of $\varepsilon$ would result in higher orders not containing all terms in their respective order as we only consider $\rho_\pm^{(c)}$ up to first order. Hence, we choose to expand the exponential to the same order as we chose for $\rho_\pm^{(c)}$. 

The elements of $\rho_\pm^{(c)}$ contain more than a factor ten more elements than in the symplectic case. Meaning that the expansion would require handling a number of terms that is in the order of tens to hundreds of thousands. Thus, we perform all further calculations with the use of \textsc{Mathematica} \cite{Mathematica} and sketch our steps.

\subsection{Integration over the Saddle Point Manifold}
\label{subsec:IntegralSaddlePoint}
First, we use \cref{eqn:rhopmelem} to explicitly determine the elements of $\rho_\pm^{(c)}$ appearing in \cref{eqn:fsdef} in the parametrization. Second, we insert the elements into \cref{eqn:cfsplit,eqn:cfnonlinearsigma} and expand in the anticommuting variables. After the expansion we integrate over all anticommuting variables which annihilates all terms not containing all anticommuting variables. Third, we integrate over the orthogonal degrees of freedom $\phi_j$ where only terms survive that were independent of $\phi_j$ since the integration of the complex exponential over a $2\pi$-periodic interval is zero. The last step is the integration over the unitary degrees of freedom. Similar to the symplectic case the expressions do not depend on the angle $\varphi_2$ and the integration is trivial. Next we integrate over $\varphi_1$ with the help of the identity
\begin{equation}
	\int_{0}^{2\pi} \dd x \exp\left(\pm i n x + c_1 e^{ix} + c_2 e^{-ix}\right) = 2 \pi I_n(2 \sqrt{c_1 c_2}) \frac{1}{\sqrt{c_1 c_2}^n} \begin{cases}
		c_2^n &, + \\
		c_1^n &, -
	\end{cases} 
\end{equation}
where $I_n(z)$ is the modified Bessel function of $n$th order. The last integral over the radial coordinate $u$ consists of Bessel functions with monomials where the sum of the order of the Bessel function and monomial is odd. The integration gives Bessel functions of varying orders which we express through Bessel functions of zeroth and first order by means of recurrence relations. Finally, we arrive at the characteristic functions for different spin orientations where
\begin{align}\label{eqn:cfGSE}
	R^{\text{GSE}}(k) =& 1 + \frac{1}{16} \int_{0}^{\infty}\limits \dd\theta \int_{0}^{\pi}\limits\dd\theta_1 \int_{0}^{\pi/2}\limits\dd\theta_2 \frac{\sin\theta_1 \sin\theta_2 \sinh^3\theta}{\left(\cos\left(\theta_1 + \theta_2\right) - \cosh\theta\right)^2 \left(\cos\left(\theta_1 - \theta_2\right) - \cosh\theta\right)^2} \notag\\
	&\times \prod_{c=1}^{M} \frac{\left(g_c + \cos\left(\theta_1 + \theta_2\right)\right)\left(g_c + \cos\left(\theta_1 - \theta_2\right)\right)}{\left(g_c + \cosh\theta\right)^2} \notag\\
	&\times k^2 \left(k^2 \left(\omega_{ab}^4 + \omega_{ab}^2 t_{ab} + t_{aa} t_{bb}\right) \frac{J_1\left(\omega_{ab} k\right)}{\omega_{ab} k} - 2 \left(2 \omega_{ab}^2 + t_{ab}\right) J_0\left(\omega_{ab} k\right)\right)
\end{align}
which was the main result from Ref. \cite{GG2025} and the spin dependent part
\begin{align}\label{eqn:cfSB}
	R_{m m\myprime}^{\text{SB}}(k) =& \int_{0}^{\infty}\limits \dd\theta \int_{0}^{\pi}\limits\dd\theta_1 \int_{0}^{\pi/2}\limits\dd\theta_2 \frac{\sin\theta_1 \sin\theta_2 \sinh^3\theta}{\left(\cos\left(\theta_1 + \theta_2\right) - \cosh\theta\right)^2 \left(\cos\left(\theta_1 - \theta_2\right) - \cosh\theta\right)^2} \notag\\
	&\times \prod_{c=1}^{M} \frac{\left(g_c + \cos\left(\theta_1 + \theta_2\right)\right)\left(g_c + \cos\left(\theta_1 - \theta_2\right)\right)}{\left(g_c + \cosh\theta\right)^2} \notag\\
	&\times \Biggl(\left((-1)^m\left(\iota_{0,ab} + \iota_{0,ab}^+ + \iota_{0,ab}^-\right) + (-1)^{m\myprime}\left(\iota_{0,ba} + \iota_{0,ba}^+ + \iota_{0,ba}^-\right) \right) J_0\left(\omega_{ab} k\right) \notag\\
	&+ \left((-1)^m\left(-\iota_{1,ab} + \iota_{1,ab}^+ + \iota_{1,ab}^-\right) + (-1)^{m\myprime} \left(- \iota_{1,ba} + \iota_{1,ba}^+ + \iota_{1,ba}^-\right)\right) \frac{J_1\left(\omega_{ab} k\right)}{\omega_{ab} k}\Biggr)
\end{align}
identifying the spin orientations $m, m\myprime=\uparrow, \downarrow$ with $0,1$, respectively. The coefficients in \cref{eqn:cfSB} are
\begin{align}\label{eqn:coefficients0}
	\iota_{0,c c\myprime} =& \frac{i}{64 \Delta \omega_{c c\myprime} \left(\cosh\theta + g_c\right)^2} \notag\\
	&\times	\Biggl(
		16 i \omega_{c c\myprime} k^2 \left(t_{c\myprime}^+ \sin\left(\theta_1 + \theta_2\right) + t_{c\myprime}^- \sin\left(\theta_1 + \theta_2\right) \right) \left(-\Delta \cosh\theta + 2 \gamma_c\right) \notag\\
		&+ 2G_{c c\myprime} \sinh\theta \left(96 + k^2 \left(4 t_{c c\myprime}^+ + 9 \omega_{c c\myprime}^2\right)\right) \left( \Delta \cos\theta_1 \cos\theta_2 + 2 \gamma_c\right) \notag\\
		&+ 8 k^2 \Delta G_{c c\myprime} t_{c c\myprime}^- \sinh\theta \sin\theta_1 \sin\theta_2 \notag\\
		&+ \Delta G_{c c\myprime} \sinh\theta \left(- 864 - k^2 \left(32 t_{c c\myprime}^+ + 45 \omega_{c c\myprime}^2\right) + k^4 \left(4 t_{c c} t_{c\myprime c\myprime} + 3 t_{c c\myprime}^+ \omega_{c c\myprime}^2 + 4 \omega_{c c\myprime}^4\right)\right)
	\Biggr) \notag\\
	\iota_{0,c c\myprime}^\pm =& \frac{1}{64 \Delta \omega_{c c\myprime} \left(\cos\left(\theta_1 \pm \theta_2\right) + g_c\right)^2} \notag\\
	&\times \Biggl(
		8 \omega_{c c\myprime} k^2 t_{c\myprime}^+ \sin\left(\theta_1 + \theta_2\right) \left(\Delta \cos\left(\theta_1 - \theta_2\right) + 2 \gamma_c\right) \notag\\
		&+ 8 \omega_{c c\myprime} k^2 t_{c\myprime}^- \sin\left(\theta_1 - \theta_2\right) \left(\Delta \cos\left(\theta_1 + \theta_2\right) + 2 \gamma_c\right) + 16 \Delta \omega_{c c\myprime} k^2 t_{c\myprime}^\pm \cosh\theta \sin\left(\theta_1 + \theta_2\right) \notag\\
		&+ i G_{c c\myprime} \sinh\theta \left(96 + k^2 \left(8 t_c^\mp t_{c\myprime}^\mp + 9 \omega_{c c\myprime}^2\right)\right) \left(\Delta \cos\left(\theta_1 + \theta_2\right) - 2 \gamma_c\right)
	\Biggr) \notag\\
\end{align}
\begin{align}\label{eqn:coefficients1}
	\iota_{1,c c\myprime} =& \frac{1}{64 \Delta \omega_{c c\myprime} \left(\cosh\theta + g_c\right)^2} \notag\\
	&\times \Biggl(
		4\omega_{c c\myprime} k^2 t_{c\myprime}^+ \left(t_c^- t_{c\myprime}^- + \omega_{c c\myprime}^2\right) \sin\left(\theta_1 + \theta_2\right) \left(\Delta \cosh\theta - 2 \gamma_c\right) \notag\\
		&+ 4\omega_{c c\myprime} k^2 t_{c\myprime}^- \left(t_c^+ t_{c\myprime}^+ + \omega_{c c\myprime}^2\right) \sin\left(\theta_1 - \theta_2\right) \left(\Delta \cosh\theta - 2 \gamma_c\right)\notag\\
		&+ 2 i G_{c c\myprime} \sinh\theta \left(192 + k^2 \left(8 t_{c c\myprime} - 38 \omega_{c c\myprime}^2\right) + k^4 \omega_{c c\myprime}^2 \left(t_{c c\myprime} + 4 \omega_{c c\myprime}^2\right)\right) \left(\Delta \cos\theta_1 \cos\theta_2 + 2 \gamma_c\right) \notag\\
		&+ 2 i k^2 \Delta G_{c c\myprime} t_{c c\myprime}^- \left( 8 + \omega_{c c\myprime} k^2\right) \sinh\theta \sin\theta_1 \sin\theta_2 \notag\\
		&+ i \Delta G_{c c\myprime} \sinh2\theta \left(-864 + k^2 \left(-32 t_{c c\myprime} +159 \omega_{c c\myprime}^2\right) + k^4 \left(t_{c c\myprime} \left(4 t_{cc} - 5 \omega_{c c\myprime}^2\right) - 20 \omega_{c c\myprime}^2\right)\right) 
	\Biggr) \notag\\
	\iota_{1,c c\myprime}^\pm =& \frac{1}{32 \Delta \omega_{c c\myprime} \left(\cos\left(\theta_1 \pm \theta_2\right) + g_c\right)^2} \notag\\
	&\times \Biggl(
		- 2 \omega_{c c\myprime} k^4 t_{c\myprime}^\pm \left(t_c^\pm t_{c\myprime}^\pm + \omega_{c c\myprime}^2\right) \sin\left(\theta_1 + \theta_2\right) \left(\Delta \cosh\theta + 2 \gamma_c\right) \notag\\
		&+ i G_{c c\myprime} \sinh\theta \left(96 + k^2 \left(t_c^\pm t_{c\myprime}^\pm \left(8 + \omega_{c c\myprime}^2 k^2\right) + \omega_{c c\myprime}^2 \left(-19 + 2 \omega_{c c\myprime}^2 k^2\right)\right)\right) \notag\\
		&\times \left(-\Delta \cos\left(\theta_1 + \theta_2\right) + 2 \gamma_c\right)
	\Biggr)
\end{align}
with
\begin{align}
	t_c^+ =& \frac{\sin\left(\theta_1 + \theta_2\right)}{\cos\left(\theta_1 + \theta_2\right) + g_c}, \quad t_c^- = \frac{\sin\left(\theta_1 - \theta_2\right)}{\cos\left(\theta_1 - \theta_2\right) + g_c}, \quad G_{c c\myprime} = \sqrt{\frac{\cosh\theta + g_c}{\cosh\theta + g_{c\myprime}}} \notag\\
	t_{cc} =& t_c^+ t_c^-, \quad t_{c c\myprime}^\pm = t_c^+ t_{c\myprime}^+ \pm t_c^- t_{c\myprime}^-, \quad \omega_{c c\myprime} = \frac{\sinh\theta}{\sqrt{\left(\cosh\theta + g_c\right)\left(\cosh\theta + g_{c\myprime}\right)}}.
\end{align}
We make a few observations. We dropped the index $s$ as it is clear from the coefficients in \cref{eqn:coefficients0,eqn:coefficients1} that the symmetry breaking part does not depend on $s$, as we already discovered for the symplectic part in Ref. \cite{GG2025}. Thus, real and imaginary part are equally distributed. This is a particularly nice result because it confirms that the deviation from the orthogonal case \cite{KNSGDMRS2013} is due to our choice of inspecting the distributions of the scattering matrix elements which hide the symplectic features of the scattering matrix itself. Hence, whether real and imaginary part are equally distributed is not a feature of the symmetries of the target Hamiltonian or the channels. Therefore, it is apparently possible to break the spin symmetry independently from breaking the degeneracy of the distributions of real and imaginary part. Furthermore, we stress that while the symmetry breaking part of the characteristic function $R_{m m\myprime}^{\text{SB}}$ differs only in relative or global signs for different spin orientations it is always measured relative to the symplectic characteristic function $R^{\text{GSE}}$. It is also important that the form of the dependence on $m, m\myprime$ is not a direct consequence of our linear approximation of the supermatrix $\rho_\pm^{(c)}$ as well as the characteristic function. Instead, higher orders display the same dependence, as directly follows from \cref{eqn:rhopm}. Further, following the reasoning from Ref. \cite{GG2025} the symmetry breaking part does not generate additional Efetov-Wegner terms in any order of $\varepsilon$ as the relevant elements of $\rho_\pm^{(c)}$ vanish at the boundaries of the integration.

It is quite remarkable that while breaking the symplectic symmetry of the coupling vectors $W_c$ we can still find an explicit expression for the characteristic function to first order in $\varepsilon$. Higher orders in $\varepsilon$ are not more complex, they only differ in the amount of terms that occur and are calculable from \cref{eqn:rhopmelem}.

\subsection{Regime of Validity for our Approximation}
\label{subsec:Validity}
We discuss the regime in which our approximation holds. We observe in \cref{eqn:coefficients0,eqn:coefficients1} that for $g_c=1$ there is a singularity at $\theta_1=\pi, \theta_2=0$. This is also the case for the expansion of $d_\pm(\varepsilon)$ in powers of $\varepsilon d_1$ as $d_1$ has the same singularity. Importantly, this divergence only arises for $g_c=1$ corresponding to perfect transmission $T_c=1$. For the sake of simplicity we set $v=1, E=0$ such that perfect transmission is equivalent to $\gamma_c=1$. This is a special case because either the upper or lower bound of $\gamma_c$ is equal to one depending on which root one chooses for a fixed $T_c$, as $g_c = -1 + 2/T_c$. If we were to choose $\gamma_c\leq 1$ then it is clear that $\varepsilon$ in \cref{eqn:wceps} is zero as otherwise one of the components would have a partial width of greater than one. The same is true for the other root, $\gamma_c\geq 1$ and also for arbitrary $v$ and appropriate energies $E$ . Hence, in the case of perfect transmission the preference of one spin orientation over the other is not possible within the scope of our proposed model which resolves the problem of divergences as we need to explicitly exclude the case in which they occur.

\section{Conclusions}
\label{sec:conclusion}
Extending our recent result for systems with symplectic symmetry, we put forward a model to study the breaking of the spin orientation symmetry in the distributions of scattering matrix elements. We introduced a relative bias between spin orientations measured by the parameter $\varepsilon$. We carried out the subsequent calculation by proper modification of the GSE case. Hereby, we encountered the necessity to only consider small parameter values $\varepsilon$ relative to the coupling strength of the channels. We presented an exact result to first order. Higher orders are calculable in the same fashion. Indeed, our results show that different spin orientations are no longer equally distributed. Furthermore, we found that while the spin symmetry is broken the real and imaginary parts of the scattering matrix elements are still equally distributed. Thus, confirming our claims in Ref. \cite{GG2025} that equally distributed real and imaginary parts are a feature of choosing the scattering matrix elements which hides the symplectic symmetry of the scattering matrix itself rather than a consequence of the symplectic nature of the system itself. As it might be of interest for experiments, we mention that it is possible to introduce a channel specific bias via a parameter $\varepsilon_c$ without much additional effort. Our results pave the way to compare experimental realizations with analytical results. Since we calculated the characteristic function, typical experimental measures such as the moments are easily obtained as derivatives of $R_{s, m m\myprime}(k)$ at $k=0$ with the help of \cref{eqn:cfGSE,eqn:cfSB}. In the case that only the cross sections are accessible it is also possible to calculate their corresponding moments from our results following the same approach as in Ref. \cite{GG2025}.

\section*{Acknowledgments}
We thank A. Aldabag, B. Dietz and S. Köhnes for fruitful discussions. This research was funded by the Deutsche Forschungsgemeinschaft (DFG, German Research Foundation) within the project Stochastic Quantum Scattering -- New Tools, New Aspects, DFG project number 540160740.
\newpage

\appendix
\section{Parametrization of the Saddle Point Manifold}
\label{app:parametrization}
The saddle point manifold is parametrized by 
\begin{align}
	Q =& - i \mathcal{U}^{-1} \begin{bNiceMatrix}
		\cos\widehat{\theta} & \sin\widehat{\theta} \\
		\sin\widehat{\theta} & - \cos\widehat{\theta}
	\end{bNiceMatrix} \mathcal{U}, \notag\\
	\widehat{\theta} =& \diag\left(\widehat{\theta}_{\text{BB}}, \widehat{\theta}_{\text{FF}}\right), \quad \widehat{\theta}_{\text{BB}} = i \theta \mathds{1}_2, \quad \widehat{\theta}_{\text{FF}} = \begin{bNiceMatrix}
		\theta_1 & \theta_2 \\
		\theta_2 & \theta_1
	\end{bNiceMatrix}
\end{align}
and the singular values fulfill $\theta\geq 0, \pi \geq \theta_1 \geq 0, \pi/2 \geq \theta_2 \geq 0$. Additionally, the transformations $\mathcal{U} = \diag\left(\widehat{U}^\dagger u_1, u_2\right)$ are
\begin{align}
	\widehat{U} =& U \oplus \mathds{1}_2, \quad U = \begin{bNiceMatrix}
		u e^{i\varphi_1} & \sqrt{1-u^2} e^{i\varphi_2} \\
		-\sqrt{1-u^2} e^{-i\varphi_2} & u e^{-i\varphi_1}
	\end{bNiceMatrix} \notag\\
	u_j =& O_j v_j, \quad O_j = \mathds{1}_2 \oplus \diag\left(e^{-i\phi_j}, e^{+i\phi_j}\right) \notag\\
	v_j =& \exp\left(i^{j-1} \left(Y - (-1)^j \frac{Y^3}{3}\right)\right), \quad Y_j = \begin{bNiceMatrix}
		0 & - \xi_j^\dagger \\
		\xi_j & 0
	\end{bNiceMatrix}, \notag\\
	\xi_j =& \begin{bNiceMatrix}
		\mu_j & \nu_j^\star \\
		\nu_j & \mu_j^\star
	\end{bNiceMatrix}
\end{align}
where $u\in[0,1], \varphi_j\in[0,2\pi], \phi_j\in[0,2\pi]$, $\mu_j, \nu_j$ and their complex conjugates are anticommuting variables and $j=1,2$. The corresponding volume element is
\begin{align}
	\dd\mu(Q) =& \mathcal{B} \dd\theta \dd\theta_1 \dd\theta_2 \dd u \dd\varphi_1 \dd\varphi_2 \dd\phi_1 \dd\phi_2 \dd[\Upsilon], \notag\\
	\mathcal{B} =& \frac{2 u \sin\theta_1 \sin\theta_2 \sinh^3\theta}{\left(\cos\left(\theta_1 + \theta_2\right) - \cosh\theta\right)^2 \left(\cos\left(\theta_1 - \theta_2\right) - \cosh\theta\right)^2}
\end{align}
where $\dd[\Upsilon]$ contains all differentials of anticommuting variables.

\section*{References}

\bibliography{bibliography.bib,streu_refs_combined.bib,Lett_Bib.bib}

\bibliographystyle{iopart-num}

\end{document}